Title: Powering population health research: Considerations for plausible and actionable effect sizes

Authors:
Ellicott C. Matthay [a,b]
Erin Hagan [a]
Laura M. Gottlieb [a]
May Lynn Tan [a]
David Vlahov [c]
Nancy Adler [a]
M. Maria Glymour [a,b]

Author affiliations:
[a] Center for Health and Community, University of California, San Francisco
3333 California St., Suite 465
Campus Box 0844
San Francisco, California 94143-0844
USA

[b] Department of Epidemiology and Biostatistics, University of California, San Francisco
550 16th Street, 2nd Floor
Campus Box 0560
San Francisco, California 94143
USA

[c] Yale School of Nursing at Yale University
400 West Campus Drive, Room 32306
Orange, CT 06477
USA

Funding: This work was supported by the Evidence for Action program of the Robert Wood Johnson Foundation (RWJF).

Role of the funding source: This work was supported by the Evidence for Action program of the Robert Wood Johnson Foundation (RWJF). RWJF had no role in the study design; collection, analysis, or interpretation of data; writing of the article; or the decision to submit it for publication.

Conflicts of interest: The authors have no competing interests to declare.

Abstract:

Evidence for Action (E4A), a signature program of the Robert Wood Johnson Foundation, funds investigator-initiated research on the impacts of social programs and policies on population health and health inequities. Across thousands of letters of intent and full proposals E4A has received since 2015, one of the most common methodological challenges faced by applicants is selecting realistic effect sizes to inform power and sample size calculations. E4A prioritizes health studies that are both (1) adequately powered to detect effect sizes that may reasonably be expected for the given intervention and (2) likely to achieve intervention effects sizes that, if demonstrated, correspond to actionable evidence for population health stakeholders. However, little guidance exists to inform the selection of effect sizes for population health research proposals. We draw on examples of five rigorously evaluated population health interventions. These examples illustrate considerations for selecting realistic and actionable effect sizes as inputs to power and sample size calculations for research proposals to study population health interventions. We show that plausible effects sizes for population health inteventions may be smaller than commonly cited guidelines suggest. Effect sizes achieved with population health interventions depend on the characteristics of the intervention, the target population, and the outcomes studied. Population health impact depends on the proportion of the population receiving the intervention. When adequately powered, even studies of interventions with small effect sizes can offer valuable evidence to inform population health if such interventions can be implemented broadly. Demonstrating the effectiveness of such interventions, however, requires large sample sizes.

Introduction

Power and sample size calculations are essential for quantitive research proposals on evaluations of population health interventions. To determine whether a proposed study is worthwhile to conduct, funders evaluate whether the study is adequately powered to detect effect sizes that may reasonably be expected for the given intervention. Thus, to ensure that studies on the impacts of population health interventions are adequately powered, researchers planning these studies must select *plausible* effect sizes as inputs to power and sample size calculations. Likely effect sizes may be estimated based on pilot studies, theories of change, causal models, expert opinion, or scientific literature on similar interventions (Leon et al., 2011; Matthay, 2020; Thabane et al., 2010). However, the relevant knowledge base for many population health interventions is sparse, which means that researchers are often only guessing at likely effects.

Evidence for Action (E4A), a Signature Program of the Robert Wood Johnson Foundation, funds investigator-initiated research on the impacts of social programs and policies to identify scalable solutions to population health problems and health inequities. Across thousands of Letters of Intent and Full Proposals E4A has received since 2015, one of the most common methodological challenges faced by applicants is predicting the likely effect size of a prospective intervention to inform power and sample size calculations. For example, of 141 invited Full Proposals, 16% (22) had reviewer concerns about anticipated effect sizes or interlocking questions about sample size and statistical power; many do not make it past the Letter of Intent stage due to power concerns. Like many funders, E4A prioritizes health studies that are adequately powered to detect effect sizes that may reasonably be expected for the given intervention. It also prioritizes intervention effects sizes that, if demonstrated, correspond to actionable evidence for population health stakeholders. General considerations for effective sample size calculations have been proposed (Lenth, 2001), but none that specifically apply to population health interventions.

In this paper, we draw on our experiences as funders of population health research, published literature, and examples of rigorously evaluated population health interventions to illustrate key considerations for selecting plausible and actionable effect sizes as inputs to power and sample size calculations. We map the reported effect estimates in our examples to standardized measures of effect to compare among them and to evaluate the relevance of established effect size benchmarks. We illustrate how to consider the impacts of the characteristics of the intervention, the mechanisms of effect, the target population, and the outcomes being studied on individual-level effect sizes achievable with population health interventions. We also use population attributable fractions, a measure of population health impact, to illustrate how various effect sizes correspond to population-level health impacts, depending on the outcome frequency and proportion of the population receiving the intervention. Although the boundaries of "population health interventions" are fuzzy, we focus here on non-medical, population-based or targeted programs or policies that are adopted at a community or higher level and affect social determinants of health or social inequalities in health.

Materials and methods

To select the examples, we reviewed population health interventions in the Community Guide (Community Preventive Services Task Force, 2019), What Works for Health consortium (County Health Rankings and Roadmaps, 2019), and Cochrane database of systematic reviews (Cochrane Library, 2019). We sought to select studies of well-established population health interventions with strong evidence on causal effects. We considered experimental and observational research, prioritizing evidence from systematic reviews, meta-analyses, or randomized trials, while recognizing that such studies are rare for population health interventions (P. A. Braveman et al., 2011). We aimed to select studies with mature evidence for which there is apparent general consensus on the intervention's health impact. We sought to select studies along a spectrum of intervention types, study population sizes, and anticipated impacts at the individual level. We sought to select a diverse set of examples that would highlight considerations for plausible effect sizes. As we reviewed the evidence, we stopped adding examples once we reached saturation with key considerations.

To compare effect sizes across studies and to evaluate the relevance of established effect size benchmarks (Cohen, 1988; Sawilowsky, 2009), we mapped the reported effect estimates in our examples (including correlation coefficients, odds ratios, relative risks, and risk differences) to standardized mean differences (SMDs, also known as Cohen's d). Box 1 presents the formulas and assumptions that we applied to convert across effect measures. Box 1 also presents the formulas we used to convert these individual effect sizes to population health impact using population attributable fractions.

For each illustrative intervention, we reviewed the existing evidence on its health effects. Interventions typically demonstrated benefits for multiple outcomes. We hypothesized that the individual-level effect sizes achieved with population health interventions would to be small relative to established effect size benchmarks. We therefore focused on the health-related outcomes with the largest effects seen for each example, because this allowed for better assessment of our hypothesis.

Results

We selected five illustrative interventions: home-visiting programs in pregnancy and early childhood, compulsory schooling laws, smoke-free air policies, mass media campaigns for tobacco prevention, and smoking cessation quitlines. Table 1 describes the content and nature of each intervention, along with the largest reported effect size across the health outcomes evaluated. Seven key considerations emerged from these examples and are described below.

*Consideration 1: Effect sizes depend on features of the intervention*

Effect sizes vary by the intervention's intensity, content, duration, and implementers. At one end are high-touch, individually-tailored interventions, typically fielded for a small number of individuals and anticipated to have lasting effects for those people directly affected; these can be considered "high-intensity". At the other end are larger, environmental interventions anticipated to have smaller individual impacts; these can be considered "low-intensity".

Home visiting programs in pregnancy and early child hood are high-touch interventions. They are individually-tailored, targeted, one-on-one interventions involving intensive supports which range in duration and can continue for multiple years. Home-visiting programs there for have larger anticipated effect sizes that compulsory schooling laws (CSLs). CSLs are a universal, low-touch, contextual intervention. They involve no individual targeting, tailoring, or person-to-person contact, and thus effect sizes are likely to be smaller. Similarly, smoke-free air policies can be considered low-touch interventions and must be enforced to be effective. Mass media campaigns to reduce tobacco use and can be considered low- to medium-touch interventions, depending on the degree of exposure. Quitlines to promote tobacco cessation are higher-touch than contextual interventions because they involve one-on-one contact with targeted individuals and usually some degree of individual tailoring, but they would still be considered light- to medium-touch, compared with home-visiting programs. Given this, selection of effect sizes should therefore be informed by the intensity of the intervention

Within these overarching intervention types, variations in the nature of the intervention drive variations in effect sizes. Effect sizes for home-visiting vary by program content and duration, and not all formats are effective (Bilukha et al., 2005; Olds et al., 2014). For example, professional home visitors are more effective than paraprofessionals, although longer durations with less-trained implementers can achieve comparable impacts, and programs with longer durations generally produce larger effects (Bilukha et al., 2005). Comprehensive smoke-free air policies and policies targeting specific industries (e.g. restaurant workers) appear to be more effective than partial bans (Community Preventive Services Task Force, 2014b; Faber et al., 2017; Frazer et al., 2016; Hahn, 2010; Hoffman & Tan, 2015; Meyers et al., 2009; Tan & Glantz, 2012). The health impacts of CSLs vary widely by setting (e.g. country, historical and political context) (Hamad et al., 2018). The most effective mass media campaigns are those with the greatest frequency, diversity, and duration of communications and that have the most graphic, emotional, or stimulating content (Bala et al., 2017; Community Preventive Services Task Force, 2016; Durkin et al., 2012). Some research suggests media campaigns must reach at least 75-80% of the target population for 1.5-2 years to reduce smoking prevalence or increase quit rates (Mozaffarian Dariush et al., 2012) or at least three years for youth campaigns (Carson et al., 2017). Three or more sessions with tobacco cessation quitlines may be more effective than single sessions (Community Preventive Services Task Force, 2014a; Fiore et al., 2008; Stead et al., 2013). Thus, when using prior literature on similar interventions to inform effect size selections, the degree to which the prospective intervention differs, for example in frequency, duration, content, or qualifications of the implementers, must be addressed to anticipate whether the expected effect sizes are likely to be smaller or larger.

*Consideration 2: Effect sizes are smaller for indirect mechanisms of effect*

Effect sizes are also smaller when the intervention impacts health through an intermediary social determinant (P. Braveman et al., 2011). Anticipating likely effect sizes requires information on both: (1) the impact of the intervention on the presumed mechanism (e.g. how much do CSLs change education) and (2) the impact of that mechanism on the outcome (e.g. how much do increases in education reduce mortality). For CSLs, each additional year of schooling was associated with a 0.03 SMD reduction in the adult mortality rate and a 0.16 SMD reduction in the lifetime risk of obesity (Hamad et al., 2018). These estimates point to the effects of education, though, not to the effects of CSLs regulating education. Given that a one-year increment in a CSL was generally associated with an average of only 0.1 additional years of schooling or less (Hamad et al., 2018), we would expect the standardized effect sizes of CSLs on mortality and obesity to be proportionally small: approximately 0.003 and 0.016, respectively. In contrast, tobacco cessation quitlines that act directly on tobacco cessation are expected to have larger effects—in this case, 0.227 SMD.

*Consideration 3: Effect sizes depend on characteristics of the target population*

Effect sizes depend on who is reached by the intervention and among whom the outcome is measured in the study (the "target population"). Interventions that serve high-need individuals and outcome measures focused on that high-need subpopulation may yield larger effects than population-level outcomes for universal interventions which affect both high- and low-need individuals. Likewise, interventions that are only relevant to a subset of the population will have larger effects when measured in that subpopulation and smaller population-level effects. Population-level interventions intended to modify determinants of health (e.g. education) instead of directly changing health are unlikely to shift those determinants for everyone. Thus, if the outcome is assessed in the overall population, the intervention effect will be an average of the null effects on people for whom education was unchanged by the intervention and the benefit for people whose education was changed by the intervention.

As evidence of this, the impacts of home-visiting programs are greater for more vulnerable families—e.g. mothers with lower psychological resources (Olds et al., 2002, 2004). Similarly, the impacts of CSLs appear to vary notably by characteristics of the recipient and changes induced by the policy, depending not just on changes in the duration of schooling but also on gender, education quality, and impacts on individuals' peers (Galama et al., 2018). Smoke-free air policies are particularly impactful when targeting particular communities or workplaces such as restaurants and bars (Community Preventive Services Task Force, 2014b; Faber et al., 2017; Frazer et al., 2016; Hahn, 2010; Hoffman & Tan, 2015; Meyers et al., 2009; Tan & Glantz, 2012). For mass media campaigns, the strongest associations are for those interventions that reach the highest proportions of the target population (Bala et al., 2017;

Community Preventive Services Task Force, 2016; Durkin et al., 2012). Campaigns may also be more successful for low-income individuals (Community Preventive Services Task Force, 2016) and for light smokers compared to heavy smokers (Secker-Walker et al., 2002). Tobacco cessation interventions are only administered to a highly selected subset of the population: tobacco-users who want to quit. Thus, the impact of quitline services will be smaller for the larger population of tobacco users, some of whom are not seeking to quit. For the same reason, sessions initiated by potential quitters may be less effective at changing population prevalence of smoking than sessions initiated by counselors (Community Preventive Services Task Force, 2014a; Fiore et al., 2008; Stead et al., 2013).

*Consideration 4: Effect sizes depend on the health outcome under study*

Effect sizes are larger for short-term, proximal outcomes compared to long-term, distal outcomes and depend on the duration of follow-up. Health behaviors such as smoking are more likely to change—and more likely to change quickly—compared to all-cause mortality. Influences on distal outcomes require longer to appear; thus, longer durations of follow-up are required and shorter follow-up periods will correspond to smaller effect sizes (one can think of 5-year mortality and 20-year mortality as two different outcomes with different likely effect sizes). For home visiting programs, although long-term impacts on distal outcomes are more difficult to realize than those on immediate outcomes such as child maltreatment episodes, high-quality implementations of the program have achieved small but nontrivial reductions in all-cause mortality among children whose mothers received home-visiting (1.6% vs 0%) 20 years after implementation (Olds et al., 2014). Similarly, CSLs may have larger effects on lifetime obesity and smaller effects on all-cause mortality (Hamad et al., 2018).

Short-term impacts—for example on smoking—may not persist, but sustained effects are possible. Studies of mass media interventions with long-term follow up suggest that effects may last for several years after program completion (Community Preventive Services Task Force, 2016). Smoke-free policies range in their effects on secondhand smoke exposure (0.54 SMD), asthma (0.13-0.17 SMD), adult tobacco use (0.09 SMD), youth tobacco use (0.09 SMD), preterm birth (0.06-0.08 SMD), hospital admissions for cardiovascular events (0.03 SMD), and low birthweight (0 SMD) (Been et al., 2014; Community Preventive Services Task Force, 2014b). Most studies assessed impacts 6-12 months post-policy change, but effects lasting up to 7 years have been documented (Community Preventive Services Task Force, 2014b). For some interventions such as quitlines, only proximal outcomes such as tobacco cessation may be feasible or realistic to collect (Community Preventive Services Task Force, 2014a; Fiore et al., 2008; Stead et al., 2013).

Beyond the long- versus short-term and distal versus proximal, population health interventions may play a different causal role, and thus have different magnitudes of effect, for different outcomes. CSLs appear to have no meaningful effect on heart disease and harmful effects on alcohol use (Hamad et al., 2018). Mass media campaigns have been more successful in preventing uptake than promoting quitting, and more influence on adult tobacco use than youth tobacco use (Community Preventive Services Task Force, 2016; Durkin et al., 2012). In this respect, theories of change and causal models may be particularly useful for evaluating the relative importance of different determinants of the outcome and thus the potential magnitude of intervention effects.

*Consideration 5: Plausible effect sizes for population health interventions may be smaller than common guidelines suggest*

Cohen's guidelines cites SMDs of 0.20, 0.50, and 0.80 as "small", "medium", and "large", respectively (Cohen, 1988). Although originally offered with many caveats, these benchmarks continued to be frequently used in research proposals, including those by E4A. Cohen's benchmarks correspond with the distribution of observed effect sizes in psychology research (M. Lipsey & Wilson, 1993; Sedlmeier & Gigerenzer, 1989), but it is unknown whether they apply to interventions related to social determinants of health. Smaller effects may be expected because population health interventions differ fundamentally from the controlled laboratory settings and short-term, proximal outcomes studied in many psychology experiments. Indeed, sociologist Rossi's Rules of Evaluation, based on years of experience evaluating social programs, emphasized that most large-scale social programs are likely to have zero net impact (Rossi, 2012).

Table 1 reports the largest effect size observed for any health outcome for each of the five illustrative interventions. Across the examples, the largest effect size was 0.54 SMD for the reduction in secondhand smoke exposure achieved by smoke-free air policies. This corresponds to a "medium" effect according to Cohen's benchmarks. The other interventions corresponded to "small" or even smaller effect sizes. Even long-term, high-intensity interventions such as home-visiting and proximal health outcomes such as smoking cessation for quitlines failed to achieve "large" effect sizes.

*Consideration 6: Translate measures of effect from prior literature to a common scale*

Likely effect sizes can be informed by existing scientific literature on similar interventions. However, determining the implications of previous studies for power calculations can be challenging because measures of effect are reported on different scales, such as the risk difference or odds ratio, and converting between scales is not always straightforward. Box 1 presents formulas that can be used to convert among effect measures, and Table 2 applies these formulas to illustrate how the magnitudes of SMDs, correlation coefficients, odds ratios, relative risks, and risk differences correspond to one another.

For the power calculations in a given research proposal, researchers will likely utilize the measures of effect that correspond to the proposed analytic strategy (e.g. odds ratios from logistic regression). The Box 1 formulas are useful to translate the effect sizes from previous studies to the scale most relevant to the research at hand. For example, if the most similar previous interventions report odds ratios for a binary outcome (e.g. poor mental health), these formulas can inform estimates of the corresponding effect size for a closely related but continuous outcome (e.g., a dimensional measure of mental health symptoms).

*Consideration 7: Small individual effect sizes can translate to large population health impact*

Beyond plausibility of anticipated effect sizes, researchers and funders must consider whether proposed studies are adequately powered to detect any effect size large enough to be important or actionable for population health. Studies should be powered to detect effect sizes that correspond to meaningful shifts in population health or health equity sufficient to justify changes in policy or practice (Durlak, 2009). Yet standardized effect sizes alone do not convey this information. Population-level effects also depend on the proportion of the population exposed to the intervention, the outcome frequency, and whether similar effect sizes can be expected in segments of the population beyond the one under study (Rothman et al., 2008). Research proposals should explicitly present the population-level effects that are likely given the anticipated effect size and proportion of the population to which the intervention could plausibly be extended. To illustrate, one way to do this is by calculating population attributable fractions (PAFs).

The PAF reflects the proportion of the negative outcome that could be averted by the given intervention. For any given intervention effect size at an individual level, the PAF can vary substantially based on the fraction of the population that is exposed and the frequency of the outcome in the unexposed. Figure 1 demonstrates this variation. In general, the PAF will be larger if the outcome is less common (because it is easier to eliminate most cases of a rare outcome than a common outcome) and larger if the exposure is more common. For example, a "medium" effect size (SMD=0.5) can correspond to a PAF of 0.01 if the outcome is common (20%) and the intervention is very selectively implemented (1%) or a PAF of 0.42, if the outcome is rare (1%) and the intervention is broadly implemented (50%).

Discussion

We provide guidance for selecting effect sizes to inform the design of adequately powered studies of population health interventions. Considering the characteristics of the intervention, the mechanisms of effect, the target population, and the outcomes being studied may help population health researchers to select more plausible effect sizes to inform power and sample size calculations. However, predicting plausible effect sizes for population health interventions is challenging even when evidence from similar interventions, theories of change, or causal models are relatively strong. In some cases, variation in the interventions we considered was the difference between a highly effective and entirely ineffective one (e.g. home-visiting programs), and in other cases, estimates across the literature were affected by the study context (e.g. CSLs). Thus, even high-quality evidence from a similar intervention may not be indicative of how a closely-related intervention will fare in a different setting (Deaton & Cartwright, 2018). This challenge may be particularly relevant for population health interventions in which the underlying mechanisms of effect are particularly complex.

The illustrative cases presented suggest that, for studies of population health interventions, researchers should anticipate smaller effect sizes to inform power and sample size calculations than Cohen's benchmarks suggest (Cohen, 1988; Sawilowsky, 2009). "Large" effect sizes, which correspond to odds or risk ratios of 4 or more, appear unlikely or exceptional. "Medium" effect sizes appear possible for (a) high-touch, long-term, intensive interventions for vulnerable populations such as high-quality home-visiting programs with low-income pregnant women; (b) proximal outcomes such as secondhand smoke exposure with smoke-free air policies; and (c) subgroups disproportionately-affected by universal interventions such as restaurant workers protected by smoke-free laws. For longer-term outcomes (e.g., 20-year mortality), more distal outcomes that were not the direct targets of intervention, and contextual interventions (e.g. compulsory schooling laws), "very small" to "small" effect sizes may be more realistic. Others have raised concerns about Cohen's benchmarks (Correll et al., 2020); downward revisions to Cohen's benchmarks in specific fields such as gerontology and personality studies may offer alternative benchmarks (Brydges, 2019; Gignac & Szodorai, 2016).

Studies of interventions with small effect sizes generally require larger sample sizes and thus more funding. Yet the typical data and funding sources available for population health intervention research often preclude the types of large-scale, high-quality studies that are necessary to definitively identify "small" or "medium" effects, even if these would be of substantial public health benefit. Larger, more appropriately powered studies could be supported by (1) more regularly collected, high-quality, individual-level, geographically-detailed administrative/surveillance data and (2) incorporating measurements of participation in population health interventions into existing large-scale primary data collection efforts (Min et al., 2019; Davis & Holly, 2006; Erdem et al., 2014).

*Actionable effect sizes for population health*

Our PAF calculations illustrate that even a very small effect size might correspond to a large population health effect if the intervention is implemented broadly. Conversely, interventions with large effect sizes may have disappointing population impacts if applied selectively. Sample size calculations can therefore also be justified using the *smallest important effect size*—i.e., the smallest effect which, if verified, would justify adoption of the intervention—because evaluating an intervention with benefits smaller than this threshold would have no actionable implications.

Every intervention entails both direct costs and opportunity costs. If the intervention is very expensive, the smallest important effect size may be large, whereas even a very small effect size might be important for an intervention that could be implemented with little cost or easily scaled up.

The biomedical, economic, social, and political considerations that affect stakeholders' evaluations of the smallest important effect size are often omitted from discussions of sample size or power. Little research exists on what PAFs are considered important or actionable for different audiences. These considerations could be amenable to quantification and potentially assessed in the same manner as power calculations when judging the rigor and importance of research proposals.

### *Limitations*

The "considerations" we present apply to quantitative, action-oriented research on the impacts of social programs and policies. Although this field is substantial in scope, different considerations may be appropriate for research in other contexts (M. W. Lipsey & Wilson, 2001). We present a small selection of examples of interventions that vary in intensity and population scope, considering both proximal and distal outcomes, to highlight key considerations for selecting realistic effect sizes for sample size and power calculations. The fact that three of these examples come from the tobacco literature reflects, to some degree, where there is greater consensus and volume of scientific literature for population health interventions. A comprehensive review of the distribution of plausible effect sizes would be valuable in future research, but the combination of small effect sizes, underpowered existing studies, and publication bias may preclude an accurate assessment. Additionally, we relied on published evaluations of interventions. Given the potential for publication bias, our estimates may over-state the plausible effect sizes.

### Conclusions

Population health researchers need realistic estimates of population health impacts to design and justify their research programs. The stakes are high: Studies designed using implausible effect sizes will lack sufficient precision to infer effects and risk concluding that an important population health intervention is ineffective. By incorporating reasonable considerations and calculations like those presented here, researchers can help to ensure that their studies are adequately powered to definitively identify important and actionable interventions for population health. Research on interventions with small individual-level effects may be critical for population health if the intervention can potentially influence a large fraction of the population. To be adequately powered, however, such research will require large sample sizes or novel linkages across large-scale datasets.

Boxes, Tables and Figures

---

**Box 1: Formulas and assumptions used to convert among measures of effect**

- Common interpretations for the standardized mean difference were drawn from Cohen (small, medium, large) (Cohen, 1988) and Sawilowsky (very small, very large, huge) (Sawilowsky, 2009).

- The standardized mean difference (SMD; Cohen's d) was defined as $d = \frac{\overline{X_1} - \overline{X_2}}{S}$, where $X_1$ and $X_2$ are the sample means in treated/exposed and untreated/unexposed groups and S is the pooled standard deviation (Borenstein et al., 2009).

- We converted from the standardized mean difference $d$ to the correlation coefficient $r$ using the formula $r = \frac{d}{\sqrt{d^2 + 4}}$. This approach assumes $r$ is based on continuous data from a bivariate normal distribution and that the two comparison groups are created by dichotomizing one of the variables (Borenstein et al., 2009).

- We converted from the standardized mean difference $d$ to the odds ratio $OR$ using the formula $OR = \exp\left(\frac{d * \pi}{\sqrt{3}}\right)$, where $\pi$ is the mathematical constant (approximately 3.14) (Hasselblad & Hedges, 1995). This approach assumes the underlying outcome measure is continuous with a logistic distribution in each exposure/treatment group.

- We converted from the odds ratio $OR$ to the relative risk $RR$ using the formula $RR = \frac{OR}{1 - P_0 + P_0 * OR}$ (Zhang & Yu, 1998), and from the relative risk $RR$ to the risk difference $RD$ using the formula $RD = P_0 * RR - P_0$, where for both, $P_0$ is the risk of the outcome in the unexposed/untreated group. For illustration, we considered a situation with a rare outcome ($P_0$=0.01) and a common outcome ($P_0$=0.20).

- Reported relative measures of association (OR, RR) that were less than 1 were inverted for comparability (e.g. an OR of 0.70 was converted, equivalently, to 1/0.70 = 1.43).

- We computed the population attributable fraction $PAF$ using the formula $PAF = \frac{P_e(RR-1)}{1 + P_e(RR-1)}$, where $P_e$ is the proportion exposed or treated and $RR$ is the relative risk (Rothman et al., 2008). For illustration, we considered $P_e$ values of 0.01, 0.20, and 0.50.

- Throughout, we assume that all measure of effect are addressing the same broad, but comparable question, and it is only the exact variables or measures that differ (Borenstein et al., 2009).

**Table 1: Characteristics and largest effect sizes in illustrative population health interventions**

| Intervention | Description | Intervention features | Target population | Largest reported effect size (SMD) | Outcome |
|---|---|---|---|---|---|
| Home visiting programs in pregnancy and early childhood | Home-visiting programs in pregnancy and early childhood are designed to provide tailored support, counseling, or training to socially vulnerable pregnant women and parents with young children. Home visitors are generally trained personnel such as nurses, social workers, or paraprofessionals. Services address child health and development, parent-child relationships, basic health care, and referral and coordination of other health and social services. Numerous variants exist, such as Healthy Families America and Nurse-Family Partnership. Programs have demonstrated benefits on a range of outcomes, including prevention of child injury, mortality, and later arrests, as well as improvements in maternal health, birth outcomes, child cognitive and social-emotional skills, parenting, and economic self-sufficiency (Bilukha et al., 2005; Office of the Surgeon General, 2001; Olds et al., 2002, 2004, 2014). | High-touch, individually-tailored, one-on-one, intensive supports, typically 1+ years in duration | Targeted to high-need individuals | 0.369 <br><br> (Bilukha et al., 2005) | Child maltreatment episodes |
| Compulsory schooling laws | Compulsory schooling laws (CSLs) increase educational attainment by requiring a minimum number of years of education among school-age children (Acemoglu & Angrist, 1999; Lleras-Muney, 2005; Hamad et al., 2018; Galama et al., 2018). CSL-related increases in educational attainment are associated with improvements in numerous health outcomes, including adult | Low-touch | Universal | 0.016 <br><br> (Hamad et al., 2018) | Obesity |

| | mortality, cognition, obesity, self-rated health, functional abilities, mental health, diabetes, and health behaviors such as smoking, nutrition, and health care utilization (Fletcher, 2015; Galama et al., 2018; Hamad et al., 2018; Ljungdahl & Bremberg, 2015; Lleras-Muney, 2005) though not all outcomes.(Hamad et al., 2018) | | | | |
|---|---|---|---|---|---|
| Smoke-free air policies | Smoke-free air policies are public laws or private sector rules that prohibit smoking in designated places. Policies can be partial or restrict smoking to designated outdoor locations. Laws may be implemented at the national, state, local, or private levels, and are often applied in concert with other tobacco use prevention interventions. There is substantial evidence that smoke-free policies have improved numerous health outcomes (Been et al., 2014; Callinan et al., 2010; Community Preventive Services Task Force, 2014b; Faber et al., 2017; Frazer et al., 2016; Hahn, 2010; Hoffman & Tan, 2015; Meyers et al., 2009; Tan & Glantz, 2012). | Low-touch | Universal or targeted to specific communities or workplaces | 0.541 (Community Preventive Services Task Force, 2014b) | Second-hand smoke exposure |
| Mass media campaigns to reduce tobacco use | Mass media interventions leverage television, radio, print media, billboards, mailings, or digital and social media to provide information and alter attitudes and behaviors. Messages are usually developed through formative testing and target specific audiences. With respect to tobacco use, campaigns have been used to improve public knowledge of the harms of tobacco use and secondhand smoke and to reduce tobacco use. Television campaigns have been most common and often involve graphic images or emotional messages. (Bala et al., 2017; Community Preventive Services Task Force, 2016; Durkin et al., 2012; | Low-touch or medium-touch, depending on exposure | Universal or targeted to key subpopulations (e.g. youth) | 0.208 | Tobacco use initiation |
| | Mozaffarian Dariush et al., 2012; Murphy-Hoefer et al., 2018) | | | | |
| Quitlines to promote tobacco cessation | Quitlines provide telephone-based counseling and support for tobacco users who would like to quit. In typical programs, trained specialists follow standardized protocols during the first call initiated by the tobacco user and several follow-up calls schedule over the course of subsequent weeks. Quitline services may be tailored to specific populations such as veterans or low-income individuals, provide approved tobacco cessation medications, or involve proactive outreach to tobacco users.(Community Preventive Services Task Force, 2014a; Fiore et al., 2008; Stead et al., 2013). | Medium-touch, sometimes individually-tailored | Targeted to current smokers who want to quit | 0.227 (Stead et al., 2013) | Tobacco cessation |

SMD: Standardized mean difference

**Table 2: Correspondence among measures of effect**

| Common interpretation | Standardized mean difference | Correlation coefficient | Odds ratio | Relative Risk | | Risk Difference | |
|---|---|---|---|---|---|---|---|
| | | | | $P_0$=0.01 | $P_0$=0.20 | $P_0$=0.01 | $P_0$=0.20 |
| Very small | 0.01 | 0.00 | 1.02 | 1.02 | 1.01 | 0.000 | 0.003 |
| - | 0.02 | 0.01 | 1.04 | 1.04 | 1.03 | 0.000 | 0.006 |
| - | 0.05 | 0.02 | 1.09 | 1.09 | 1.07 | 0.001 | 0.015 |
| - | 0.10 | 0.05 | 1.20 | 1.20 | 1.15 | 0.002 | 0.031 |
| - | 0.15 | 0.07 | 1.31 | 1.31 | 1.24 | 0.003 | 0.047 |
| Small | 0.2 | 0.10 | 1.44 | 1.43 | 1.32 | 0.004 | 0.064 |
| - | 0.3 | 0.15 | 1.72 | 1.71 | 1.51 | 0.007 | 0.101 |
| - | 0.4 | 0.20 | 2.07 | 2.04 | 1.70 | 0.010 | 0.141 |
| Medium | 0.5 | 0.24 | 2.48 | 2.44 | 1.91 | 0.014 | 0.182 |
| - | 0.6 | 0.29 | 2.97 | 2.91 | 2.13 | 0.019 | 0.226 |
| - | 0.7 | 0.33 | 3.56 | 3.47 | 2.35 | 0.025 | 0.271 |
| Large | 0.8 | 0.37 | 4.27 | 4.13 | 2.58 | 0.031 | 0.316 |
| - | 0.9 | 0.41 | 5.12 | 4.91 | 2.81 | 0.039 | 0.361 |
| - | 1 | 0.45 | 6.13 | 5.83 | 3.03 | 0.048 | 0.405 |
| - | 1.1 | 0.48 | 7.35 | 6.91 | 3.24 | 0.059 | 0.448 |
| Very large | 1.2 | 0.51 | 8.82 | 8.18 | 3.44 | 0.072 | 0.488 |
| - | 1.3 | 0.54 | 10.57 | 9.65 | 3.63 | 0.086 | 0.525 |
| - | 1.4 | 0.57 | 12.67 | 11.35 | 3.80 | 0.103 | 0.560 |
| - | 1.5 | 0.60 | 15.19 | 13.30 | 3.96 | 0.123 | 0.592 |
| - | 1.75 | 0.66 | 23.91 | 19.45 | 4.28 | 0.185 | 0.657 |
| Huge | 2 | 0.71 | 37.62 | 27.54 | 4.52 | 0.265 | 0.704 |
| - | 2.25 | 0.75 | 59.21 | 37.42 | 4.68 | 0.364 | 0.737 |
| - | 2.5 | 0.78 | 93.18 | 48.48 | 4.79 | 0.475 | 0.759 |

See Box 1 for formulas and assumptions used to convert among measures of effect. $P_0$: Risk of outcome among unexposed or untreated.

**Figure 1: Population attributable fractions for varying effect sizes (SMD), baseline risks ($P_0$), and proportions exposed ($P_e$)**

| Common interpretation | Standardized mean difference (SMD) | $P_0$ | Population attributable fraction | | |
|---|---|---|---|---|---|
| | | | $P_e$=0.01 | $P_e$=0.20 | $P_e$=0.50 |
| Very small | 0.01 | 0.01 | 0.00 | 0.00 | 0.01 |
| | | 0.2 | 0.00 | 0.00 | 0.01 |
| Small | 0.2 | 0.01 | 0.00 | 0.08 | 0.18 |
| | | 0.2 | 0.00 | 0.06 | 0.14 |
| Medium | 0.5 | 0.01 | 0.01 | 0.22 | 0.42 |
| | | 0.2 | 0.01 | 0.15 | 0.31 |
| Large | 0.8 | 0.01 | 0.03 | 0.39 | 0.61 |
| | | 0.2 | 0.02 | 0.24 | 0.44 |
| Very large | 1.2 | 0.01 | 0.07 | 0.59 | 0.78 |
| | | 0.2 | 0.02 | 0.33 | 0.55 |
| Huge | 2 | 0.01 | 0.21 | 0.84 | 0.93 |
| | | 0.2 | 0.03 | 0.41 | 0.64 |

$P_0$: Risk of the outcome among the unexposed. $P_e$: Proportion of the population exposed. Values in the shaded cells are population attributable fractions. "Common interpretation"s are based on Cohen's benchmarks (Cohen, 1988).